\documentclass[3p]{elsarticle}

\usepackage[UKenglish]{babel}
\usepackage{lineno}
\usepackage{hyperref}
\usepackage{float}
\usepackage{graphicx}
\usepackage{xcolor} %for textcolor
\usepackage{bm}
\usepackage{amsmath, amssymb, amsfonts, amsbsy}
\usepackage[normalem]{ulem}
\usepackage{upgreek}

\modulolinenumbers[10]

\journal{npj Computational Materials}

\hypersetup{
    colorlinks=true,
    linkcolor=blue,
    urlcolor=blue,
    citecolor=red
}

\usepackage{outlines}
\usepackage[colorinlistoftodos,prependcaption,textsize=footnotesize]{todonotes}

\usepackage[normalem]{ulem}

\begin{document}

%\setmainfont{Times New Roman} 

\begin{frontmatter}

\title{Deciphering Acoustic Emission with Machine Learning}
%\author{Dénes Berta, Balduin Katzer, Katrin Schulz, Péter Dusán Ispánovity}
\author[ELTEaddress]{D\'{e}nes Berta}
\author[KITaddress,HKAaddress]{Balduin Katzer}
\author[KITaddress,HKAaddress]{Katrin Schulz\corref{mycorr}}\cortext[mycorr]{Corresponding authors}
\ead{katrin.schulz@kit.edu}
\author[ELTEaddress]{P\'{e}ter Dus\'{a}n Isp\'{a}novity\corref{mycorr}}\ead{ispanovity.peter@ttk.elte.hu}

\address[ELTEaddress]{ELTE E\"{o}tv\"{o}s Lor\'{a}nd University, Department of Materials Physics, P\'{a}zm\'{a}ny P\'{e}ter s\'{e}tany 1/a, 1117 Budapest, Hungary}
\address[KITaddress]{Karlsruhe Institute of Technology, Institute for Applied Materials (IAM), Kaiserstr. 12, 76131 Karlsruhe, Germany}
\address[HKAaddress]{Hochschule Karlsruhe - University of Applied Sciences (HKA), Moltkestr. 30, 76133 Karlsruhe, Germany}

\begin{abstract}
Acoustic emission signals have been shown to accompany avalanche-like events in materials, such as dislocation avalanches in crystalline solids, collapse of voids in porous matter or domain wall movement in ferroics. The data provided by acoustic emission measurements is tremendously rich, but it is rather challenging to precisely connect it to the characteristics of the triggering avalanche. In our work we propose a machine learning based method with which one can infer microscopic details of dislocation avalanches in micropillar compression tests from merely acoustic emission data. As it is demonstrated in the paper, this approach is suitable for the prediction of the force-time response as it can provide outstanding prediction for the temporal location of avalanches and can also predict the magnitude of individual deformation events. Various descriptors (including frequency dependent and independent ones) are utilised in our machine learning approach and their importance in the prediction is analysed. The transferability of the method to other specimen sizes is also demonstrated and the possible application in more generic settings is discussed.
\end{abstract}


\end{frontmatter}

\linenumbers
 
\section*{Introduction}

It was shown that at the micron-scale and below crystalline materials (as well as many other heterogeneous materials) exhibit complex deformation behaviour including size-related hardening and significant sample-to-sample variation in the plastic response \cite{fleck1994strain, uchic2004sample, uchic2009plasticity}. In addition, in this regime deformation becomes intermittent and constitutes of a series of random strain bursts that make the details of the deformation process unpredictable both in time and space \cite{dimiduk2006scale, csikor2007dislocation}. This intermittency and stochasticity originates from the sudden rearrangement events of the dislocation network, the so-called \emph{dislocation avalanches}.
%It is important to emphasise that although dislocation avalanches are most prominent in micron-scale samples, they are ubiquitous in crystalline plasticity, but in macroscopic samples the events are relatively small. \pdicomment{Shall we keep this sentence?}
Similar avalanche-like processes, such as the plastic deformation of amorphous materials \cite{falk1998dynamics, talamali2011avalanches, sandfeld2015avalanches, cao2018nanomechanics, tyukodi2019avalanches}, the porous collapse \cite{nataf2014avalanches, nataf2014predicting, kun2014rupture, baro2018experimental,  chen2019acoustic}, the domain wall dynamics in ferroics \cite{zapperi1998dynamics, repain2004creep, herranen2019barkhausen, casals2020avalanches, shao2022acoustic}, the stick-slip behaviour of granular matter \cite{nasuno1998time, lherminier2019continuously} or the fracture of paper \cite{salminen2002acoustic, rosti2010statistics}, are ubiquitous in nature and have gained massive interest in the last few decades as well as the universal features of avalanches \cite{laurson2013evolution}. In order to experimentally study the underlying physical process, that is, the source of the avalanche, it has to be connected to directly related proxies that are experimentally measurable with sufficient precision. This is usually a rather challenging task, since avalanches are fast and mainly occur inside the material below its surface.

Dislocation avalanches were shown to be accompanied by strong burst-like acoustic emission (AE) signals which are elastic waves triggered by the release of stored elastic energy at the onset of the event \cite{weiss1997acoustic, weiss2003three, miguel2001intermittent, weiss2015mild}. As such, AE is a natural candidate as the proxy of the avalanche activity in crystals. It was also demonstrated that other deformation mechanisms, e.g., twinning \cite{vinogradov2016limits, vinogradov2016acoustic, chen2021real, vctvrtlik2022plastic, toth2023scaling} or martensitic transformation \cite{vives1995statistics, vives2011imaging, mathis2011investigation, planes2013acoustic, toth2014calorimetric, toth2016simultaneous, hauvsild2010characterization} have similar concomitant AE signals. However, when it comes to evaluating the measured AE data, one has to face fundamental problems that stem from the complex nature of the involved physical processes. Firstly, the acoustic waves can be subjects of significant distortion during their travel in the specimen due to reflection and interference, thus, their final form upon reaching the detector depends on the location of their source, the geometry of the sample, the quality of the contact between the sample and detector, etc. Secondly, resonance occurs in the piezoelectric transducer that is unavoidably affected by its properties (geometry, material characteristics, etc.). Finally, the necessary strong electric amplification is influenced by the features of the data acquisition hardware and software. The whole process can be mathematically formulated as $A_\mathrm{AE}(t) = V(t) \ast T(t)$, where $A_\mathrm{AE}(t)$ is the measured AE signal, $V(t)$ is the source function (representing, e.g., the local deformation rate during an avalanche) which is convolved with a $T(t)$ transfer function, that depends on all the specific details mentioned above \cite{casals2021duration, bronstein2024uncovering}. The fundamental question of AE measurements is, whether and to what extent it is possible to infer $V(t)$ from $A_\mathrm{AE}(t)$, that is, how one should interpret the local deformation event from the measured AE signal.

A very common approach to investigate avalanches through the lens of AE is the analysis of some easily accessible features (such as, amplitude, duration, rise time, etc.) of signals \cite{mathis2012exploring, vcapek2014study, lebedkina2018crossover,chen2019acoustic,salje2019ferroelectric,ali2019observations,casals2021duration} which relies on a usually threshold-based, somewhat arbitrary AE event detection. Yet, these features do show clear connection with the properties of the avalanche source. A prominent example is the energy associated with AE events that was found to follow a robust power-law distribution for dislocation avalanches, a universal feature that was also recovered from a number of independent experiments, e.g., \cite{weiss1997acoustic, 
richeton2005dislocation, weiss2007evidence,lherminier2019continuously,  ispanovity2022dislocation,chen2024avalanche}, and this was also confirmed by micromechanical experiments on other types of deformation processes and simulations \cite{hidalgo2002evolution, dimiduk2006scale, csikor2007dislocation, zaiser2008strain, cui2016controlling, ispanovity2022dislocation, ugi2023irradiation}. But the rich AE signal (due to its unparalleled time resolution, usually recorded at MHz or higher) may contain far more information about the source than a few noise-affected and threshold-dependent parameters. The question naturally arises: What information (related to the particular avalanche source) is hidden in the acoustic data and how could it be deciphered?

In this work exactly this issue is addressed by analysing experiments where both $A_\mathrm{AE}(t)$ and $V(t)$ is accessible. Two current set-ups exist that offer this possibility: (i) micropillar compression coupled with AE and carried out in situ in a scanning electron microscope (SEM) where local deformation can be detected both using the microdeformation stage and also visually from the SEM images \cite{ispanovity2022dislocation, ugi2023irradiation}. (ii) And more recently Bronstein et al.~designed a specific experiment where AE detection was coupled with acceleration measurements to reveal the avalanche source during twinning in Mg samples of cm size range \cite{bronstein2024uncovering}. Here we focus on the first method, in which the whole jerky stress-strain curve is available with clear signatures (stress drops) of avalanche activity. We address the general question posed above by defining a more specific, technologically rather relevant task: Whether and to what extent can one reconstruct the plastic response of a specimen merely from acoustic data?

\begin{figure}[ht!]
    \centering
    \includegraphics[width=\textwidth]{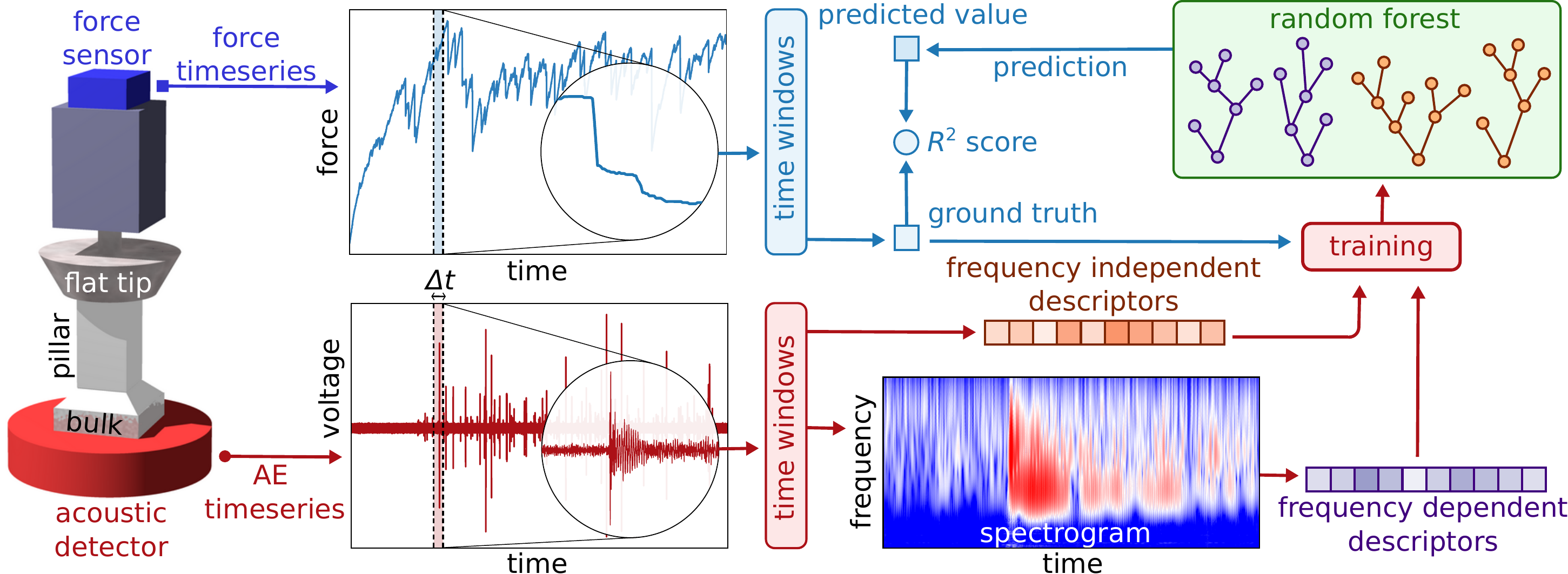}
    \caption{\textbf{The workflow of the ML approach.} Micropillar compression tests provide force and AE data which are, then, divided into equisized time windows. From the windows, frequency independent and dependent descriptors and force increments (ground truth) are extracted. These are utilised for the training of random forest ML models.}
    \label{fig:abstract}
\end{figure}

To answer this question, a set of micropillar compression tests are utilised during which single crystalline Zn micropillars of various sizes (8,16 and 32~$\upmu$m) and oriented for basal slip are compressed in situ in an SEM and AE is measured simultaneously \cite{ispanovity2022dislocation, ugi2023irradiation}. The sketch of the experimental set-up is shown in the left panel of Fig.~\ref{fig:abstract}. The micropillars (strongly magnified in the schematic) are carved in a macroscopic sample that is firmly attached to the AE sensor. The pillars are compressed using a flat punch diamond tip, that is attached to a force sensor. The latter is based on an elastic spring whose elongation is measured to obtain the acting force. During a measurement the top end of the spring is moved with constant velocity and no feedback for stress or strain control is applied. In this set-up strain bursts cause the elongation of the spring that are measured as sudden force drops (for a recorded jerky force-time curve see Fig.~\ref{fig:abstract}). For more details on the experimental set-up, see the Methods section and Ref.~\cite{ispanovity2022dislocation} This set of experiments is ideal for our purpose for two reasons. Firstly, in the given material and orientation plastic deformation is solely caused by dislocation slip on the basal plane, and no other deformation mechanism (such as twinning or crack formation) is active. Secondly, because of the absence of forest dislocations (and any other relevant short-range interaction that would hinder dislocation motion and  localise avalanches\cite{berta2023dynamic, berta2024avalanche, berta2025identifying}) dislocation avalanches are large and, thus, produce strong, easily measurable AE signals \cite{weiss2015mild} (for an example see Fig.~\ref{fig:abstract}). To sum up, in these experiments both the force and the acoustic data are available making the prediction of the former based on the latter feasible.

Machine learning (ML) has emerged as a powerful tool for extracting relevant information from huge datasets. The richness of our experimental AE data makes application of ML feasible The feasibility is demonstrated by the successful utilisation of ML for other AE-related tasks (e.g., for quality monitoring of samples) in the past \cite{RouetLeduc2017,shevchik2018acoustic, shevchik2019deep}. For this task the random forest (RF) regressor is employed which is more straightforward to interpret compared to more complex neural network based methods. This enables us to analyse the importance of different features of the AE data in the prediction, as it will be shown in the paper. In order to carry out this prediction task two approaches are introduced in this paper. Both methods rely on the simultaneously measured force and AE data for the training of the ML model which is then utilised to predict the force-time response. For the visualisation of the workflow see Fig.~\ref{fig:abstract}. The advantage of our approaches is that they do not rely on arbitrary thresholding that necessarily loses the information about acoustic signals below a certain size but are based on general descriptors of the AE timeseries such as its statistical moments (in the first approach) or features of its spectrograms (in the second approach).

The main objective of the paper is, therefore, to demonstrate the predictability of the plastic response based on merely the recorded AE signal. The potential foreseen application of such a finding would be deciphering AE data for bulk samples, where no concurrent measurements on plastic events are available. For this reason strong attention will be devoted to the question whether an ML model trained on samples of a certain size can make reasonable predictions for other specimens of unseen sizes.

\section*{Results}

\subsection*{Relevant time scales of the experiment and the concept for prediction}

In order to reveal the connection between the deformation processes and the AE activity, the workflow summarised in Fig.~\ref{fig:abstract} is utilised to predict the force-time response merely from the AE data. As seen, descriptors are defined based on AE timeseries in equisized time windows. Before choosing the window width let us first investigate the relevant time scales of the experiment. (1)~On a fine time scale the force-time curve is characterised by strong fluctuations due to force drops caused by dislocation avalanches. This scale can be related to the duration of the deformation processes (directly measurable via the timescale of the stress drops, see Suppl.~Fig.~\ref{SM:fig:duration}), typically less than 0.3~s. A division below this timescale could lead to events split into several windows, which could potentially lead to performance loss in the prediction. (2)~On a coarser scale the force-time curve is still fluctuating with alternating regimes of mainly strain hardening and softening. This scale can be roughly associated with typical waiting times between the force drops corresponding to duration of the quiescent periods between deformation events. These intervals can last up to $\sim$100~s (see Suppl.~Fig.~\ref{SM:fig:waiting_time}). It is noted that this timescale is inversely proportional with the compression rate (or, more precisely, platen velocity) \cite{ispanovity2022dislocation}.

The ML workflow should be able to reproduce both features, that is, the details of force drops on the fine scale as well as the overall shape of the force-time curve. To this end, predictions are made on both mentioned timescales. (1)~On the fine scale the force change $\Delta F$ between the beginning and end of the time window is predicted based on descriptors calculated from the AE signal. The corresponding time window width $\Delta t$ is determined based on Suppl.~Fig.~\ref{SM:fig:duration} and is chosen as $\Delta t = 300$~ms. (2) On the coarse scale the force~$F$ itself is predicted at the end of each time window of duration $\Delta T$. For the coarse width $\Delta T=50$~s was chosen based on Suppl.~Fig.~\ref{SM:fig:waiting_time}). The combination of the two predictions leads to a final prediction which inherits both the fine fluctuations and the coarse shape of the curve (for more details on how the two predictions are combined, see the Methods section).

\subsection*{Machine learning model and feature selection}

\label{sec:ML}

\begin{figure}[H]
    \centering
    \includegraphics[width=\textwidth]{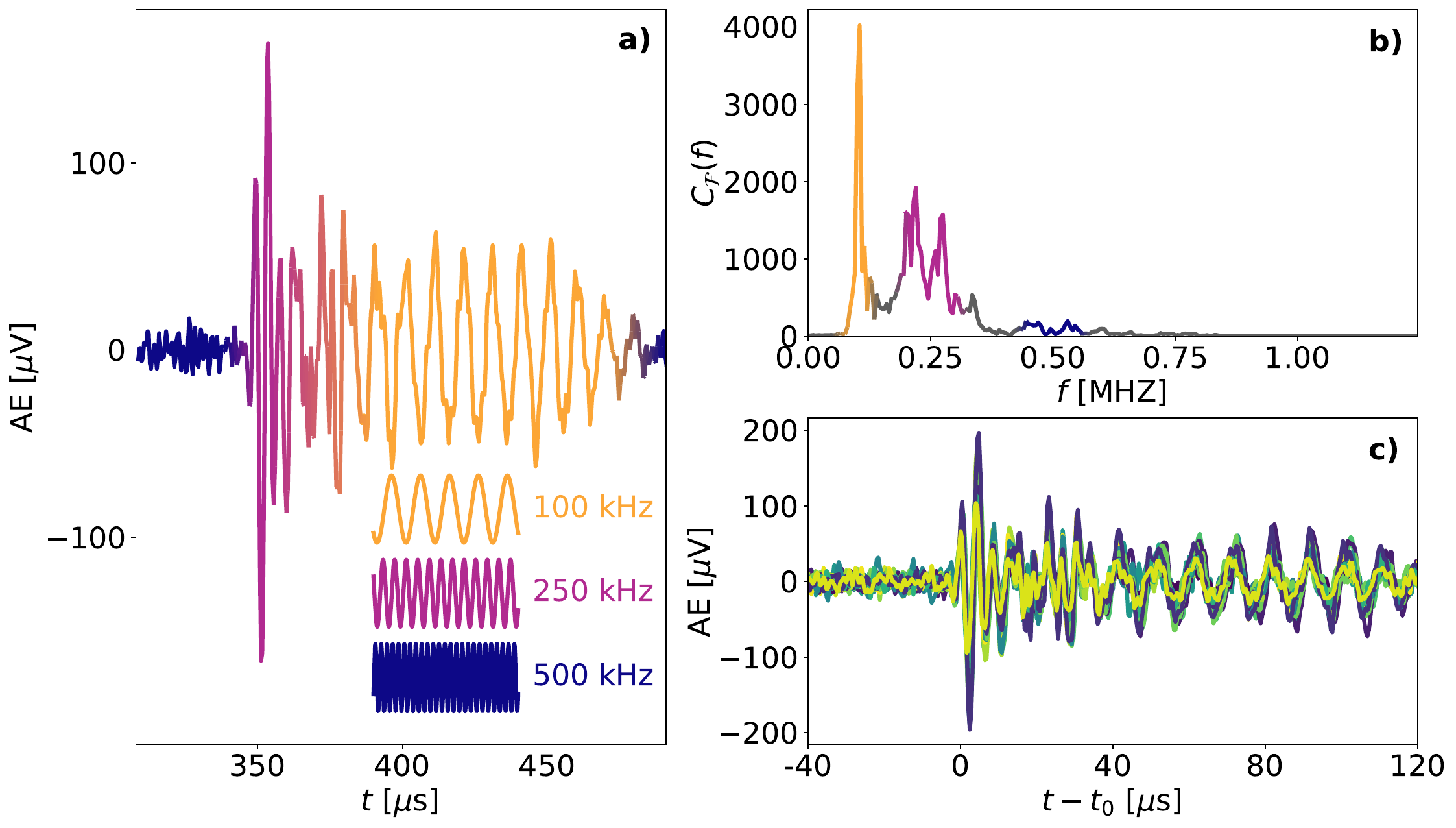}
    \caption{\textbf{Acoustic signals from compression tests of micropillars with a diameter of $\pmb{8\ \upmu}$m.} (a): A representative AE signal that constitutes of a few characteristic frequencies. The noise have a mean frequency of $\sim$500~kHz. When an event occurs a short ($\sim$20~$\upmu$s long) signal with a frequency of $\sim$250 kHz emerges from the noise which is followed by a longer periodic pattern of frequency $\sim$100~kHz. (b): The absolute value of the Fourier transform $C_\mathcal{F}$ of the frequency $f$ averaged on several time windows centred on signals. The colour-coding is consistent with the legend of panel a) and it highlights the peaks corresponding to the characteristic frequencies. (c): Typical signals shifted with $t_0$ such that their starting points coincide at $t-t_0=0$. These curves demonstrate that the signals and the involved frequencies are rather similar despite their different amplitude/energy.}
    \label{fig:frequencies}
\end{figure}

In order to predict the force increment (on the fine scale) or the force value (on the coarse scale) for the time windows, the AE activity within the respective time window has to be represented properly. To this end, firstly, the nature of the AE data at hand should be understood. Figure~\ref{fig:frequencies}(a) shows a typical AE burst during the deformation of a $8~\upmu\mathrm{m}$ sample. At the beginning of the event, a medium-frequency ($\sim$250~kHz) pattern emerges from the high-frequency ($\sim$500~kHz) background noise. This is followed by an oscillation with somewhat lower amplitude and frequency ($\sim$100~kHz), presumably due to the mechanical resonance occurring in the piezoelectric transducer. The Fourier analysis of the ensemble of signals detected in $8~\upmu\mathrm{m}$ micropillars shows that there are indeed characteristic frequency ranges (around the three specific values mentioned above) indicated by peaks in the spectrum [see Fig.~\ref{fig:frequencies}(b)]. These characteristic frequency ranges correspond to the three major regimes observed, that is, to the background noise, the initial response and the resonant part. This results in a strong similarity among acoustic signals as demonstrated by Fig.~\ref{fig:frequencies}(c). This observation motivates us to focus on information related to these specific frequencies when predicting plastic response based on AE.  Therefore, two different approaches are applied: one which does not and one which does contain frequency information about the raw AE data. Accordingly, \emph{frequency independent} and \emph{frequency dependent} features are utilised to represent the AE information in these two cases.

In the first approach statistical features of the AE timeseries (corresponding to time windows)
are computed and used as input descriptors. The descriptors are normalised moments of form $\sqrt[k]{\langle |V|^k\rangle}$.  Here, $V(t)$ is the raw AE signal and $\langle\bullet\rangle$ is the average w.r.t.~time within the relevant time window. Moments with low integer values of $k$ (such as the mean and the standard deviation) are applied as well as the maximum of $V$ (which is the $k\xrightarrow{}\infty$ limiting case). These moments encode information about the fluctuations in the AE signal, however, these descriptors completely lack frequency information.

The second approach is based on the spectrograms  of the AE time windows obtained by wavelet transformation (for specific details see the Methods section). The transformation represents the timeseries in the 2D time $\otimes$ frequency domain.  
After obtaining the spectrogram of the time windows, one-dimensional slices are taken corresponding to certain frequencies. Then, moments (of the same form as in the first approach) are computed from this 1D data. This procedure leads to a larger number of features (if the same amount of moments are used) and a richer input for the ML model since it also contains frequency information.

\begin{figure}[H]
    \centering
    \includegraphics[width=0.75\textwidth]{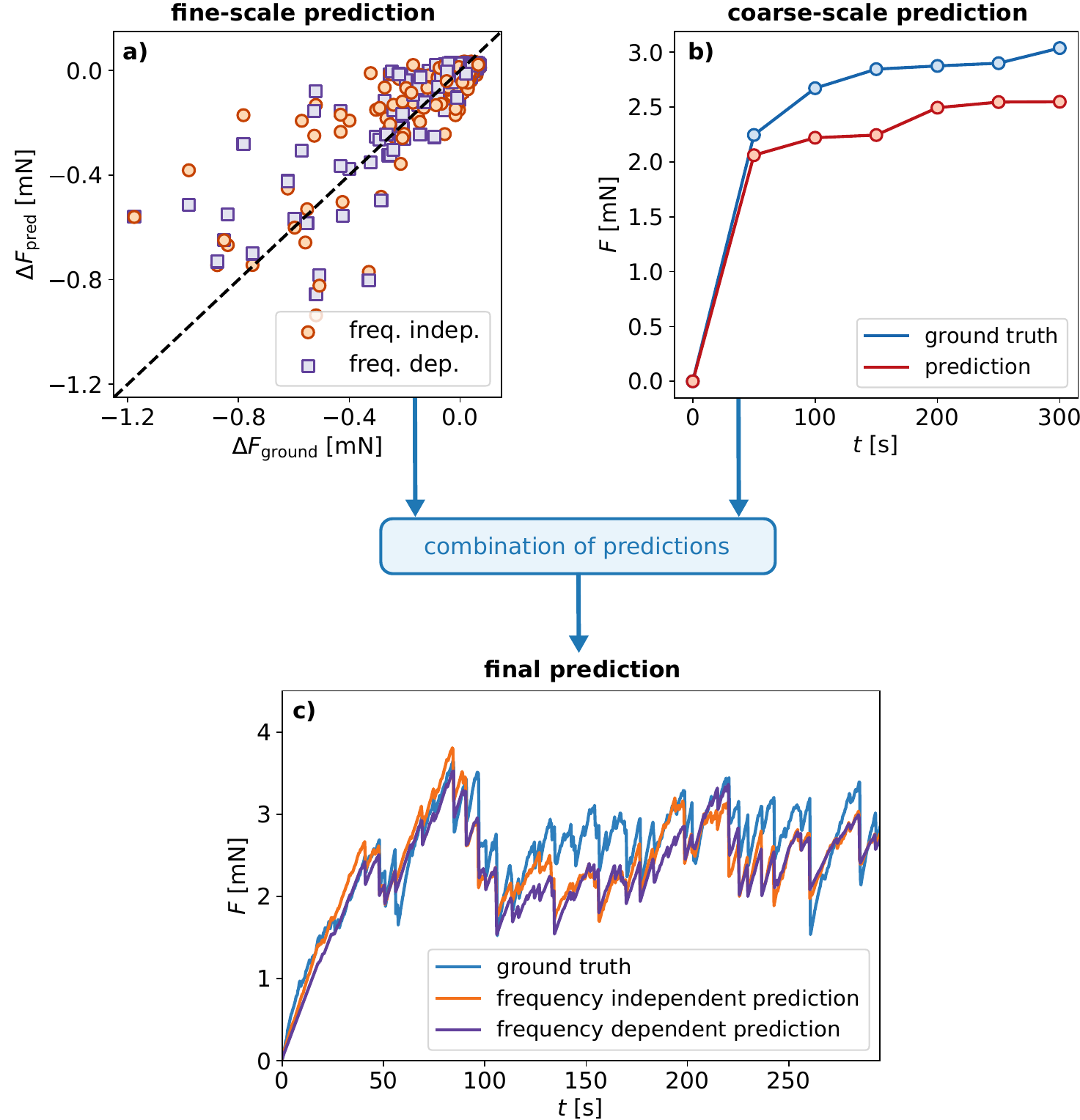}
    \caption{\textbf{Double-scale prediction of the force-time response of micropillars.} (a): Predictions of fine-scale response of a micropillar of diameter of $8~\upmu\mathrm{m}$ using frequency independent features. The data points correspond to force increments in $\Delta t = 300$~ms wide time windows. $\Delta F_\mathrm{ground}$ and $\Delta F_\mathrm{pred}$ are the ground truth and the prediction of the force increment, respectively. (b): Prediction (using frequency independent features) of the coarse-scale behaviour of the same micropillar with a time-resolution of 50~s.  (c): Comparison of the final prediction (obtained by combining the fine- and coarse-scale prediction) with the ground truth for an 8-micron specimen.}
    \label{fig:predictions}
\end{figure}

Figure~\ref{fig:predictions} shows the course of the prediction process for an arbitrarily selected example experiment conducted on a micropillar of diameter $d=8\upmu$m. The scatter plot of Fig.~\ref{fig:predictions}(a)  compares the actual (ground truth) force increments $\Delta F_\mathrm{ground}$ at the fine scale of $\Delta t = 300$~ms and the corresponding predicted values $\Delta F_\mathrm{pred}$ obtained with both the frequency independent and dependent prediction. The apparent correlation of the actual and the predicted force increment values indicate that the fine-scale prediction is able to uncover when the force drops occur based on the acoustic data. Supplemental~Fig.~\ref{SM:fig:windows_width} indicates that the performance score $R^2$ (for its definition see the Methods section) is quite robust to the variation of the window width $\Delta t$ above the duration of the longest force drops [$\sim$$200-300$~ms, see Suppl.~Fig.~\ref{SM:fig:duration}(c)]. As it can be expected, however, if some force drops are longer than the window size, the $R^2$ drops, that is, the prediction becomes worse for $\Delta t \to 0$. These observations validate the choice of $\Delta t=300~$ms which already falls into the regime where the performance scores are high and robust. It can be also noted that the frequency dependent approach achieves appreciably higher scores implying that the frequency information is valuable for the ML model. The coarse shape of the force-time curve can also be predicted well by the coarse-scale prediction scheme. It is demonstrated by the closeness of the predicted force values to the ground truth ones resulting in a predicted curve shape very similar to the ground truth as shown in Fig.~\ref{fig:predictions}(b). The final predications are obtained by the combination of the predictions of the two scales (according to the procedure presented in the Methods section). Figure~\ref{fig:predictions}(c) shows that the predicted force-time curve (for both approaches) matches very well the ground truth curve. In particular, the overall shape of the curves are very similar on the coarse time scale and remarkable accordance is seen in terms of the fine serrations between the ground truth and the predictions. The comparison of the predictions (of both approaches) and the ground truth curves for all the individual experiments are presented in Suppl.~Fig.~\ref{fig_sup:all_pred}.

\begin{figure}[ht!]
    \centering
    \includegraphics[width=0.95\linewidth]{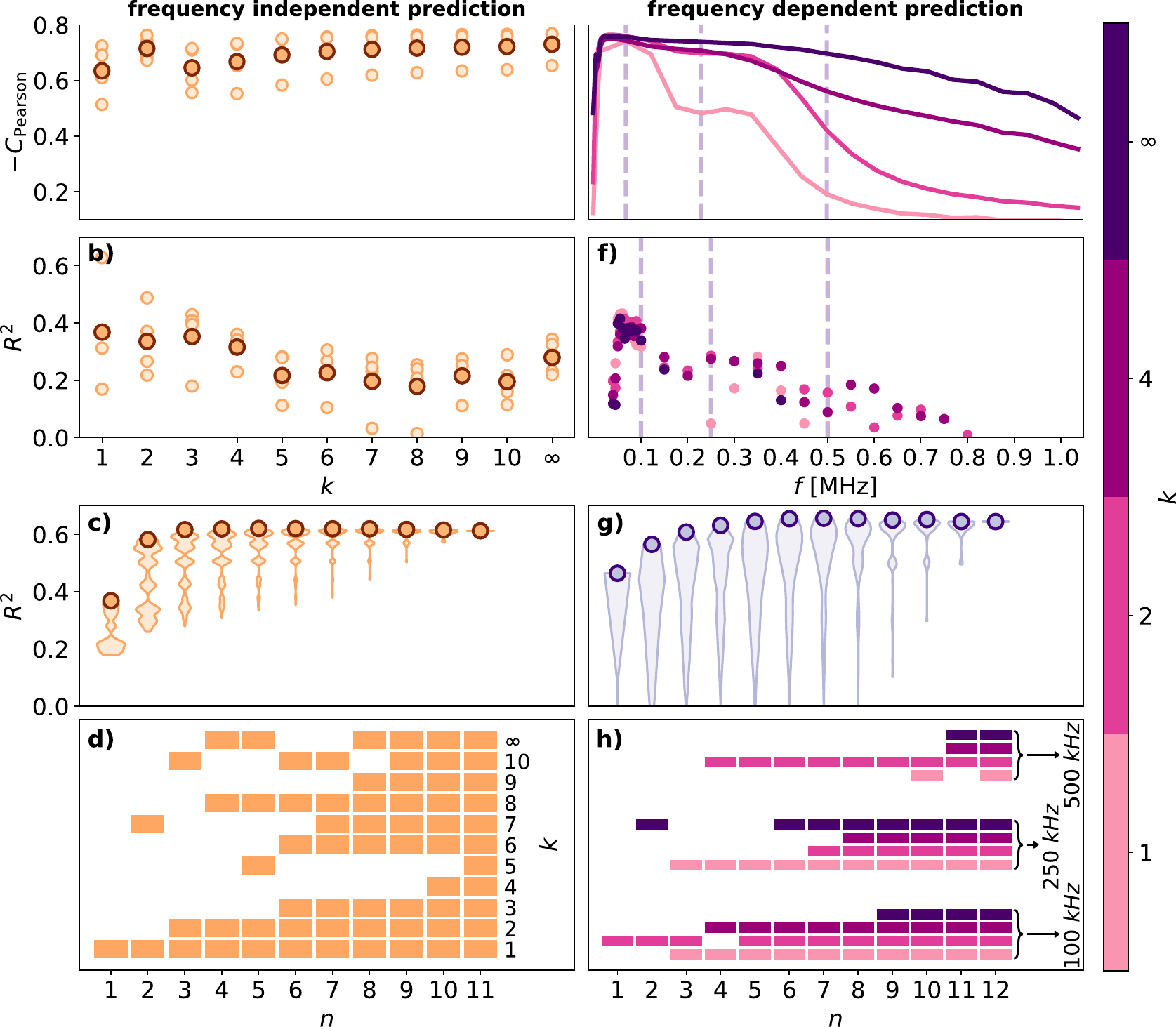}
    \caption{\textbf{Feature importance for $\pmb{d=8\upmu}$m samples.} (a): The Pearson correlation of the frequency independent features (of form $\sqrt[k]{\left\langle|V|^k\right\rangle}$ where $V$ is the raw (voltage) signal from the piezo-sensors with the target variable (for increment) for individual 8-micron experiments (small markers) and the mean for the four experiments (large markers). $k\xrightarrow {}\infty$ corresponds to the maximum. (b): The $R^2$ score for prediction based on single features for with the same meaning of small and large markers as in (a). The features best-correlated with the target variable are typically the high-order moments, however, the best scores can be achieved using lower moments. (c): The $R^2$ score for prediction based on subsets of $n$ features. All combinations are shown (light point clouds) and the best scores are highlighted with the darker markers. The best score already saturates at around $n=3$. The subset of features corresponding to the best score for each $n$ is presented in (d). Clearly, the lowest moment is of the utmost importance, however, the subset performing the best usually combines lower and higher moments as well. (e): The Pearson correlation $C_\mathrm{Pearson}$ of the frequency dependent features with the target variable (force increment) averaged for all 8-micron experiments. Different colours correspond to features based on different order moments (mean, standard deviation, the fourth root of the kurtosis and maximum) of slices of the spectrograms of the AE data (see colourbar). (f): The mean $R^2$ score for predictions based on single features. The correlations and scores clearly show that information of lower frequencies (100-300 kHz, especially around 100 kHz) that are characteristic to acoustic events is the most important. (g): The $R^2$ score for prediction based on subsets of $n$ features. The features used are the moments of order $k=1,2,4,\infty$ at frequencies 100~kHz, 250~kHz and 500~kHz [marked with dashed lines in panel (e) and (f)].  All combinations are shown (violin plots) and the best scores are highlighted with the markers. The best score already saturates at around $n=4$. (h): The subset of features corresponding to the best score for each $n$. The optimal model compositions suggest that the combined use of 100 kHz and 250 kHz features is very beneficial.}
    \label{fig:feature_importance}
\end{figure}

In the following, the importance of the features (of both approaches) in the fine-scale prediction is analysed in Fig.~\ref{fig:feature_importance}. Specifically, their correlation is examined as well as the performance (in terms of $R^2$ scores) of models trained on single features or on combinations of features. In the case of the frequency independent approach the following conclusions can be drawn. The Pearson correlation of the moments of type $\sqrt[k]{\langle |V|^k\rangle}$ and the target variable (that is, the force increment) exhibits a non-trivial dependence on $k$ saturating at a were strong anti-correlation for $k\xrightarrow[]{}\infty$ [see Fig.~\ref{fig:feature_importance}(a)]. Interestingly, when trained on single moments, the performance of our ML model shows a very different (but also non-monotonic) trend indicating that both the lower moments (such as the mean and the standard deviation) and the maximum contain a comparable amount of useful information [see Fig.~\ref{fig:feature_importance}(b)]. The analysis of $R^2$ scores of models trained on subsets of moments reveal that the prediction quality quickly saturates (already at around 3 descriptors) if the number of features is increased and the descriptors are chosen correctly [see Fig.~\ref{fig:feature_importance}(c)]. That is, even with the utilisation of a few moments the optimal predictive power of our approach can be reached. Figure~\ref{fig:feature_importance}(d) shows the optimal choice of features for given number of features $n$. The constitution of the ideal descriptor choices imply that while the lowest moments are typically very important, in order to achieve optimal performance, lower and higher moments have to be combined.

Similar conclusions can be drawn on the feature importance for the second approach. In this case moments (of order $k$) were computed based on the slices of the spectrograms corresponding to a varied frequency $f$. The analysis of Pearson correlations and single-feature $R^2$ scores highlights the importance of characteristic frequencies also observable for the individual the AE signals (see Fig.~\ref{fig:frequencies}). Both the correlations and the scores (Fig.~\ref{fig:feature_importance}(e) and Fig.~\ref{fig:feature_importance}(f), respectively) accentuate that the most important frequency range is around 100~kHz (characteristic to the low-frequency tail of the AE signals) and the higher frequencies (up to $\simeq 350$~kHz) corresponding to the first spikes of the signal also contain valuable information. The predictive power of very low and very high frequencies decays quickly. In order to test the synergy of descriptors four different moments were computed for three different frequencies: 100~kHz (low-frequency tail of signals), 250~kHz (high-frequency burst at signal onset) and 500~kHz (noise). Figure~\ref{fig:feature_importance}(g) demonstrates that, similarly to the frequency-independent case, the best $R^2$ score can be closely approached with the utilisation of very few (around 4) appropriately chosen features. The constitution of the best set of features (for a given number of features) confirms the superior importance of low-frequency features already implied earlier and that it is beneficial to combine low-frequency descriptors with middle-frequency ($\simeq$250~kHz) ones [see Fig.~\ref{fig:feature_importance}(h)].

From the definition of the features utilised, it is evident that two of them are closely related to traditional measures for characterizing acoustic bursts. Firstly, the moment-type feature of $k=2$ is related to the time-integrated $|V|^2$ representing the \emph{dissipated energy} associated with the acoustic event. Secondly, the case $k=\infty$ representing the maximum of $|V|$ is known as the \emph{amplitude} of the acoustic burst. These two characteristics of the acoustic signals are widely-used to quantify their properties \cite{casals2021duration, chen2021real, chen2022multiple}. While the amount of dissipated energy and the amplitude have been frequent choices as burst features due to their physical interpretability, our results shed light on their key importance in the prediction. Figure~\ref{fig:feature_importance}(d) shows that each optimal feature set of a size of $n>2$ contains the energy-related $k=2$ feature and either the  amplitude ($k=\infty$) or the $k=10$ feature strongly correlated (with a Pearson correlation coefficient of 0.995) to the amplitude. The importance of the energy and the amplitude is also accentuated by Fig.~\ref{fig:feature_importance}(h). The figure reveals that the most important energy-type feature is the one corresponding to frequency 100~kHz which is understandable since the 100~kHz regime (which takes significantly longer than the initial 250~kHz oscillation, see Fig.~\ref{fig:frequencies}) of a burst corresponds most to the energy type feature. Out of the amplitude-type ($k=\infty$) features, the most important is naturally the one corresponding to 250~kHz since this is the frequency corresponding to the high-amplitude initial oscillations (see Fig.~\ref{fig:frequencies}). It is particularly remarkable that these are exactly the two features ($k=2$ at 100~kHz and $k=\infty$ at 250~kHz) that constitute the best performing $n=2$ representation of the AE signal. Thus, it can be concluded that the often used energy and amplitude descriptors (combined with ML) can used to recover information about the deformation events very well, however, as Fig.~\ref{fig:feature_importance} demonstrates, the inclusion of further features of the AE response is beneficial and can significantly improve our prediction further. It is also noted, that the lowest-order (that is, $k=1$) moment-type feature is proportional to the characteristic termed as the measured area under the rectified signal
envelope (MARSE) \cite{tang2017pattern, ali2019observations, stepanova2019acoustic, stepanova2021acoustic}. This descriptor of the AE burst is less frequently used (due to the lack of simple physical interpretation), however, its immense importance in the prediction [see Fig.~\ref{fig:feature_importance}(d)] suggests that it is closely related to the physical source (dislocation avalanche in our case) of the burst.

\subsection*{Transferability of the method}

\begin{figure}[H]
     \centering
     \includegraphics[width=\textwidth]{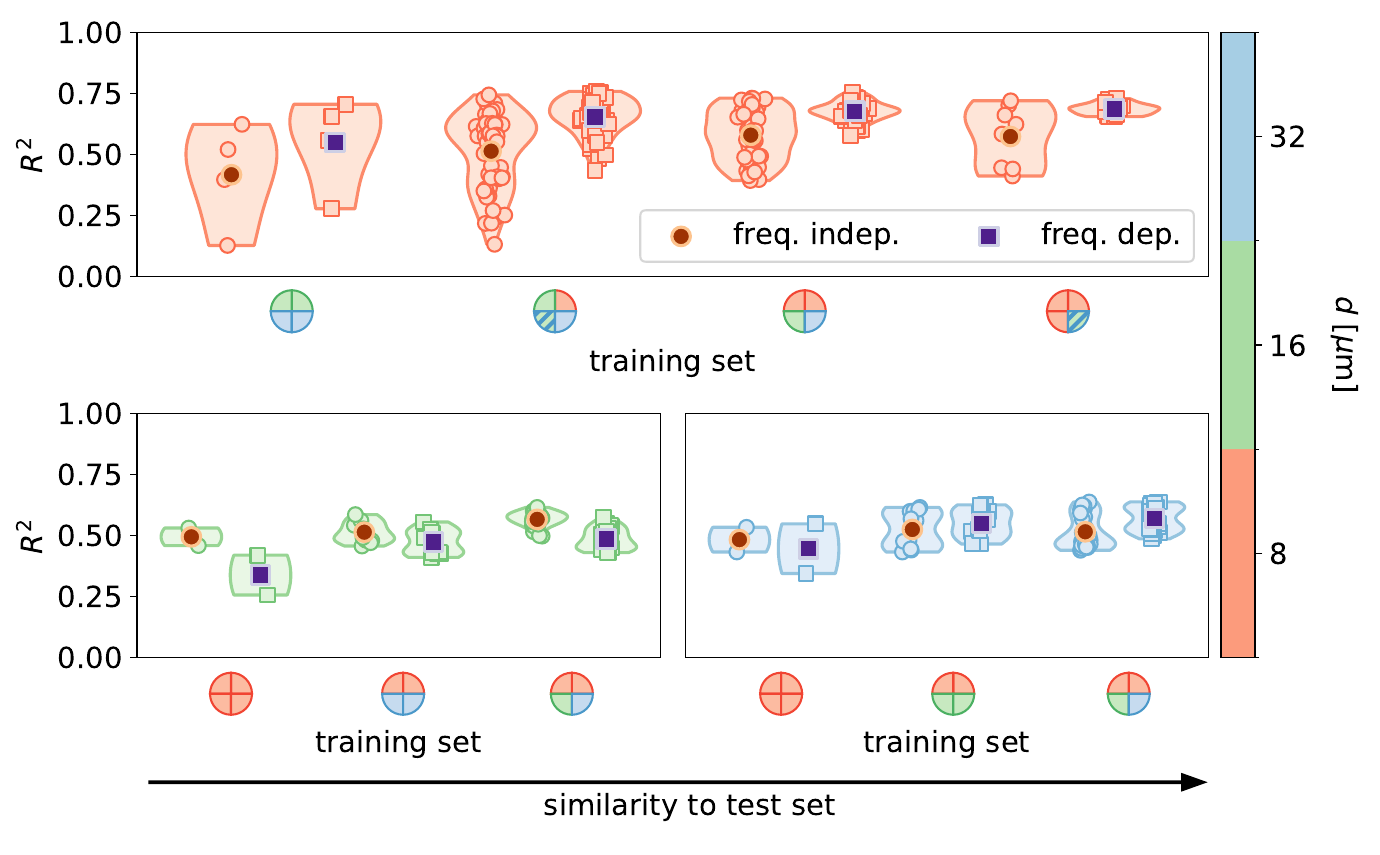}
     \caption{\textbf{Transferability of the ML method across different sample sizes.} The $R^2$ score for different combinations of 4-experiment of training sets and 1-experiment test sets. The combinations with the test set of the same micropillar diameter are arranged in the same panel the main colour of the panel indicating the diameter corresponding to the test set. The composition of the training sets are showed by the miniature pie charts. The striped slices denote a training experiment that is either of one diameter or of another. In each scenario two point clouds and violin plots indicate the performance for the frequency independent and frequency dependent approaches. The mean $R^2$ scores are indicated by the orange and purple markers in the two cases. Generally, it can be observed that, even though, the prediction score get better as the similarity of the training and the testing data increases, the model performs reasonably well even if it was not shown experiments of the same pillar size as in the test experiment. This observation holds regardless of the inclusion/exclusion of frequency information.
     }
     \label{fig:extrapolation}
\end{figure}

The capability of predicting the plastic response merely from the acoustic emission activity was demonstrated above for 8-$\upmu$m specimens (see Fig.~\ref{fig:feature_importance}). In that setup, prior to the prediction, the ML model was trained on experiments with the same characteristics (same material, same pillar size, etc.) as the test experiment. For the possible applications, however, it is also important whether it is possible to transfer the trained model to other samples where the direct determination of details of the plastic activity is impossible or cumbersome (for instance, in the case of bulk specimens).
In order to test the transferability of our method across different sample sizes the following approach is utilised. Sets of experiments are created all consisting of four experiments but with varying composition in terms of pillar diameter ($8, 16$ and $32 \, \upmu$m). Our ML model is trained on these sets separately and then is tested on a fifth different experiment. This is examined for different training and test set compositions (see Fig.~\ref{fig:extrapolation}).

The obtained results (quantified with $R^2$ scores) are summarised in Fig.~\ref{fig:extrapolation}. As it can be expected, the scores indicate that the model performs better (on average) if experiments of the same type are included in the training set (or if the ratio of this type is increased). It is remarkable, however, that the scores only moderately lower when the training and validation sets are disjoint in terms of pillar diameter. That is, the model can extrapolate (either to smaller or larger scales) and interpolate to pillars that are of previously unseen type. These results are promising in regards with the potential application of the method to  unseen specimen sizes (possibly even to bulk samples).

\section*{Discussion and Conclusion}

The AE technique is widely utilized for detecting and monitoring damage progression in various structures. It has gained recognition in the engineering community as one of the most reliable and well-established methods in non-destructive testing (NDT). AE technology has proved highly efficient and effective for identifying fracture behaviour and fatigue in materials such as metals, fiberglass, wood, composites, ceramics, concrete, and plastics \cite{gholizadeh2015review}. However, as it was elucidated in the introduction, AE measurements are inherently characterized by a significant loss of information due to various physical and technological reasons. This loss of information is encompassed in the transfer function. Yet, the success of the various AE applications means that AE data still contains significant and relevant information on damage progression. The facts that AE datasets are huge due to the large sampling rate and that relevant information is obscured in a noisy and fluctuating signal calls for the application of advanced data science methods. Indeed, analysis of AE data is an ideal application for ML, as shown by the recent sheer increase in its applications including damage mechanism identification in composites \cite{muir2021damage, almeida2023identifying}, crack classification \cite{das2019machine, ju2022machine}, leakage detection \cite{banjara2020machine, ullah2023pipeline}, life time prediction of bearings \cite{elforjani2017prognosis}, real-time quality monitoring in additive manufacturing \cite{shevchik2018acoustic, shevchik2019deep} and welding \cite{asif2022machine}. All these applications have in common, that the source of the AE signals (fiber debonding, crack propagation, etc.) cannot be directly observed, so the AE analysis is inherently phenomenological and mainly focuses on classification of events and or detection of undesired behaviour. In other words, these methods are unable to characterize the actual transfer function and, consequently, to assess properties of the individual micromechanical damage events. 

ML applications on AE employ various ML models to capture the information hidden in acoustic signals. The approaches include ones based on traditional ML algorithms such as random forest \cite{morizet2016classification, shevchik2016prediction, RouetLeduc2017, wang2020acoustic, ferrando2021novel, ullah2023pipeline}, support vector machine \cite{elforjani2017prognosis, das2019machine, banjara2020machine} or k-nearest neighbors \cite{pandya2013fault, ech2017unsupervised, das2019machine, muir2021damage, almeida2023identifying, ullah2023pipeline} and more complex, neural network based approaches based on artificial neural networks \cite{de2008health, kalafat2015acoustic, elforjani2017prognosis, jierula2021study, ullah2023pipeline}, convolutional neural networks \cite{xia2017fault, shevchik2018acoustic, ebrahimkhanlou2018single, shevchik2019deep, hasan2019acoustic, konig2021machine, sikdar2022acoustic} and recurrent neural networks \cite{marchi2017deep, zheng2018automatic, li2019gear, konig2021data}. For these ML applications the acoustic data is typically represented by conventional signal characteristics (such as duration or peak frequency) \cite{banjara2020machine,muir2021damage, ju2022machine, asif2022machine,ren2023recognition, almeida2023identifying}, statistical moments of the timeseries \cite{RouetLeduc2017,johnson2021laboratory, kononenko2023situ, ullah2023pipeline} or by spectrograms \cite{shevchik2018acoustic, ebrahimkhanlou2018single, shevchik2019deep, tran2020drill, muir2021damage}. The latter two approaches were employed in our two ML methods. Our method fits into this zoo of different strategies by applying a random forest approach to data represented by statistical moments and features extracted from spectrograms. The choice of ML model was motivated by the small number of individual experiments which results in relatively low variety in the data. A simpler model such as the random forest typically outperforms more data-hungry neural networks in such a scenario. If, however, significantly larger number of experiments were available, more complex neural network based approaches would be expected to take advantage of the rich AE data better. It is also noted the similarly to this paper our works also demonstrated the importance lower moments (e.g., the variance) in capturing the relevant information in AE data which could be used to, e.g., predict the time to failure in in sheared fault gouge materials \cite{RouetLeduc2017}. This suggests that representing the AE timeseries by its moments (especially by the lower ones) may be a good choice in general.

Our work focused on a case when dislocation slip was basically the sole deformation mechanism and we investigated the deformation of microsamples. In this case we were able to decipher the AE waveforms of individual AE events in the sense, that we could predict the corresponding plastic strain increments with high fidelity and also the general shape of the force-time curve. 
Although our results correspond to a specific case of dislocation mediated plasticity, we believe that these results represent a new direction in the utilisation of acoustic data. Namely, in an analogous experimental set up where there is a different dominant deformation mechanism (e.g., twinning or fracture), our methodology has a great potential to recognise and quantify those deformation events. Collecting such data of different deformation mechanisms could, then, lead to a more general model that can detect and classify AE events originating from different, coexisting deformation mechanisms based not on implicit assumptions but verified training data provided by in-situ experiments. This model would, at the same time, quantify the individual plastic events that are the sources of the AE events, which would represent a great leap in NDT methodologies both from the theoretical and technological perspective. 

\section*{Methods}

\subsection*{Experiment and data preparation}
\noindent
Micromechanical compression tests of single crystalline zinc micropillars oriented for basal slip were conducted at room temperature while recording both the force and the acoustic emission induced by dislocation avalanches.
The micropillars had a square-shaped cross-section and an aspect ratio of height to side of 3:1. The diameter of micropillars was either 8\,\textmu m, 16\,\textmu m or 32\,\textmu m in the experiments. Each compression test was performed with a platen velocity of 10~nm/s and a spring constant of 10~mN/$\upmu$m of a customized nanoindenter~\cite{Hegyi2017}. Indentation depth and load was monitored with an accuracy of $\sim$1 nm and $\sim$1~$\upmu$N, respectively. The force was recorded with a sampling rate of 200~Hz. An acoustic emission detector of Physical Acoustics Corporation~(PAC) with a wide-band (100–1000 kHz) was mechanically bonded on the bottom of the micropillar while ensuring a constant contact pressure and recorded acoustic data with a sampling rate of 2.5~MHz. It is important to note, that all the micropillars were milled into the same sample and the experiments were conducted during one measurement event, that is, the sample was not detached from the AE transducer between the tests and the same conditions/settings applied to all of the experiments. 
More detailed information about the experimental setup of the micropillar compression tests such as the acoustic emission detector, the amplifier and the data acquisition system, can be found in Ispánovity et al.~\cite{ispanovity2022dislocation}.

Before the extraction of features from the AE data and apply machine learning methods, the recorded AE data of each experiment was normalised by the mean of its absolute values.
The AE and force data of each experiment was subdivided into equisized time windows $\Delta t$.
Subsequently, various statistical frequency independent and frequency dependent features were extracted from each window of the acoustic emission data (as described in the main text) and the force increment as well as the force value was extracted from each window of the force data.

\subsection*{Wavelet Transformation}
\noindent
Wavelet transformation was conducted to decompose the AE signal into a localised frequency-time domain.
Formally, a wavelet transformation \(W(a, b)\) at scale \(a\)~(dilatation) and position \(b\)~(translation) is  expressed as
\begin{equation}
    W(a, b) = \frac{1}{w(a)} \int_{-\infty}^{\infty} x(t) \cdot \psi\left(\frac{t - b}{a}\right) \, dt
\end{equation}
where \(x(t)\) is the signal of time~$t$, \(\psi(t)\) is the wavelet function and \(w(a)\) is a weighting function, which is set to \(1/\sqrt{a}\) in this work to ensure energy conservation~\cite{Addison2017}.
In this work, we applied  a continous wavelet transformation~(CWT) based on the third-order Gaussian wavelet function, which is defined as
\begin{equation}
    \psi(t) = \frac{\partial^3}{\partial t^3}C e^{-t^2}=-4Ct(2t^2-3)e^{-t^2}
    \label{eq:mother_wavelet}
\end{equation}
where $C$ is a normalisation constant.
This wavelet was chosen as wavelet function due to its asymmetry and its shape fitting well to our detected AE signals~(cp.~Suppl.~Fig.~\ref{SM:fig:examplewavelet}).
In this work, we employ the \textit{PyWavelets} implementation (\emph{gaus3}) of the third-order Gaussian wavelet~\cite{Lee2019} for the CWT of the AE.
Further information on wavelets and wavelet transformations is provided in Supplementary note 2.

For applying the CWT, we generated a custom set of scales~$a$, which inversely relates to a set of frequencies~$f$ characteristic to the wavelet function.
The power spectra of the AE signals of different specimen sizes are shown in Supplementary note 3.
Thus, we chose $a$ based on a frequency range from 50~kHz to 800~kHz to incorporate all characteristic frequencies of dislocation avalanches.

To create a time-frequency dependent power spectrum, which leads to a so-called spectrogram, each AE time window~$w$ is transformed by the CWT.
An example of an AE signal as well as the corresponding spectrograms during a regime without any detected dislocation motion and during a dislocation avalanche are presented in Supplementary note 4. After the creation of spectrograms, they are utilised to extract frequency dependent descriptors as described in the main text.

\subsection*{Machine Learning at the fine-scale}
\noindent
To predict the force increment for each time window based on its extracted features, the regression model called Random Forest (RF) regressor was utilised with the \textit{scikit-learn} python implementation.
The train-test split was executed by using a set of experiments as the training data and one experiment as the test data.
For the analysis of the transferability capability of the transfer learning approach (see Fig.~\ref{fig:extrapolation}), we additionally ensured that the amount of training data remains constant throughout all permutations. This was achieved by keeping the number of considered time windows $n_{w}$ constant (based on the shortest experiment).
Based on the constant $n_{w}$, only the last $n_{w}$ time windows for each experiment were included for training.
This leads to a train:test ratio of 80\%:20\% for the transfer learning approach.
Additionally for the transfer learning approach, to reduce the number of features of the wavelet-based approach, only the three most important frequencies 100~kHz, 250~kHz and 500~kHz were included.
Hyperparameters of the frequency independent and the frequency dependent approaches were evaluated based on a grid search five-fold cross-validation and were frozen after finding the best set.
For measuring the accuracy of the regression model, we applied the coefficient of determination $R^2$ to the predicted and the actual force increments. This measure is defined as
\begin{equation}
    R^2(\bm y,\bm {y^\mathrm{(p)}})=1-\frac{\sum_{i=1}^N{\left(y_i-y_i^\mathrm{(p)}\right)^2}}{\sum_{i=1}^N{\left(y_i-\overline{y}\right)^2}}
\end{equation}
where $\bm y$ and $\bm {y^\mathrm{(p)}}$ are the vectors of the ground truth and predicted values of the target variable (in our case, the force increment), respectively, $N$ denotes the number of data point in the test set and $\overline{y}=\frac{1}{N}\sum_{i=1}^N{y_i}$. 

\subsection*{Machine Learning at the coarse-scale}
\noindent
At the coarse-scale the force value is predicted at certain times instead of force increments. For this prediction, additional features in combination with the aforementioned frequency independent features (i.e., moments of the AE timeseries) used at the fine-scale are utilized. Firstly, the number of AE signals is counted (based on thresholding), and secondly, the total duration of these AE signals is detected, i.e. features representing the number and the length of dislocation avalanches occurring in a coarse time window are used.
To ensure that the size effect is captured, the pillar diameter is added as a feature at the coarse-scale. The window width chosen is $\Delta T=50~\mathrm{s}$. Due to the large window width overlapping windows are used for training to avoid insufficient amount of training data. Adjacent windows are shifted from each other by $\Delta T/10=5~\mathrm{s}$. To predict the shape of the force-time curve of a certain experiment at the coarse-scale all these coarse-scale windows of all other experiments are utilised to train a random forest regressor model. An example of the final coarse-scale prediction is shown in Fig.~\ref{fig:predictions}(b) capturing the overall shape of the force-time curve.

\subsection*{Combination of fine- and coarse-scale prediction}
\noindent
In order to keep the fine-scale force drop structure while obtaining force-time curve that has approximately correct coarse-scale shape, the fine- and coarse-scale predictions are combined as follows. Corrections are made to the fine-scale results in each coarse window subsequently starting with the first coarse window. Let $F$ denote the coarse-scale force prediction and $f$ the force prediction obtained by summing up the force increments originating from the fine-scale prediction. Let us now consider the first coarse window. At the end of this window the two predictions provide $F(\Delta T)$ and $f(\Delta T)$. To correct the shape of the $f-t$ force-time curve we add the slope of 
\begin{equation}
    \delta_1=\frac{F(\Delta T)-f(\Delta T)}{\Delta T}
\end{equation}
to the fine-scale prediction in this first coarse window. This results in a corrected fine-scale prediction of
\begin{equation}
    f_1(t)=f(t)+\delta_1 t.
\end{equation}
Then, we move on the the second coarse windows and execute the correction similarly, then, for the third, etc. The correction of the fine-scale prediction in the $n$th coarse window is done based on the following formula 
\begin{equation}
    f_n(t)=f_{n-1}(t)+\delta_n [t-(n-1)\Delta T],\quad(\Delta T(n-1)<t)
\end{equation}
where
\begin{equation}
    \delta_n=\frac{F(n\Delta T)-f_{n-1}(n\Delta T)}{\Delta T}.
\end{equation}
Let us denote the number of coarse windows with $N_\mathrm{c}$. After iterating through all the $N_\mathrm{c}$ coarse windows, we get our final prediction of $f_N(t)$. 

This method ensures that the predicted stress--time curve intersects the ($t$,$\sigma$) points predicted by the coarse-scale prediction by adding a constant value to the force increments obtained by fine-scale prediction. This values is constant within a coarse time windows but can vary window to window. Since this procedure only adds a constant slope to the curve within a coarse time window, the curve is adjusted to follow the coarse-scale prediction while the relative differences between the increments on the fine-scale are kept unchanged within a coarse window. Consequently, the method leaves the fine structures revealed by the fine-scale prediction unaffected.

\section*{Data availability}
The experimental data used in this study was originally published from Isp\'{a}novity et al.~\cite{ispanovity2022dislocation} and is deposited in the Zenodo database at \url{https://zenodo.org/records/5897653}.
The processed data that support the findings of this study are available from the corresponding authors upon reasonable request.

\section*{Author contribution}
These authors contributed equally: D\'{e}nes Berta, Balduin Katzer. P.D.I. and K.S. designed the research and supervised the project. D.B. and B.K. processed the experimental data, developed the ML model and analysed and visualized the results. D.B., B.K., K.S. and P.D.I. wrote the paper, with contributions from all authors.

\section*{Acknowledgements}
The authors would like to thank D.~Ugi for fruitful discussions and for providing experimental data. D.B. and P.D.I acknowledge financial support from the National Research, Development and Innovation Fund of Hungary under the young researchers’ excellence program NKFIH-FK-138975, from the Ministry of Innovation and Technology of Hungary from the National Research, Development and Innovation Fund, financed under the ELTE TKP 2021-NKTA-62 funding scheme, and from the European Union project RRF-2.3.1-21-2022-00004 within the framework of the MILAB Artificial Intelligence National Laboratory. B.K. and K.S gratefully acknowledge the financial support for this work in the context of the DFG research project SCHU~3074/4-1 and the funding by the Carl-Zeiss-Stiftung. This work was performed on the HoreKa supercomputer funded by the Ministry of Science, Research and the Arts Baden-Württemberg and by the Federal Ministry of Education and Research, Germany.

\section*{Competing interests}
The authors declare no competing interests.

\bibliographystyle{naturemag}
\bibliography{AE}

% SUPPLEMENT
\renewcommand\thesection{Supplementary note \arabic{section}}

\renewcommand{\figurename}{Supplementary~Figure}
\renewcommand{\thefigure}{S\arabic{figure}}
\renewcommand{\tablename}{Supplementary~Table}
\renewcommand{\thetable}{S\arabic{table}}
\setcounter{figure}{0} 
\setcounter{section}{0} 

\section*{Supplementary note 1: Relevant time scales}
\label{sec_sup:ForceIncrement}

\begin{figure}[H]
    \centering
    \includegraphics[width=\textwidth]{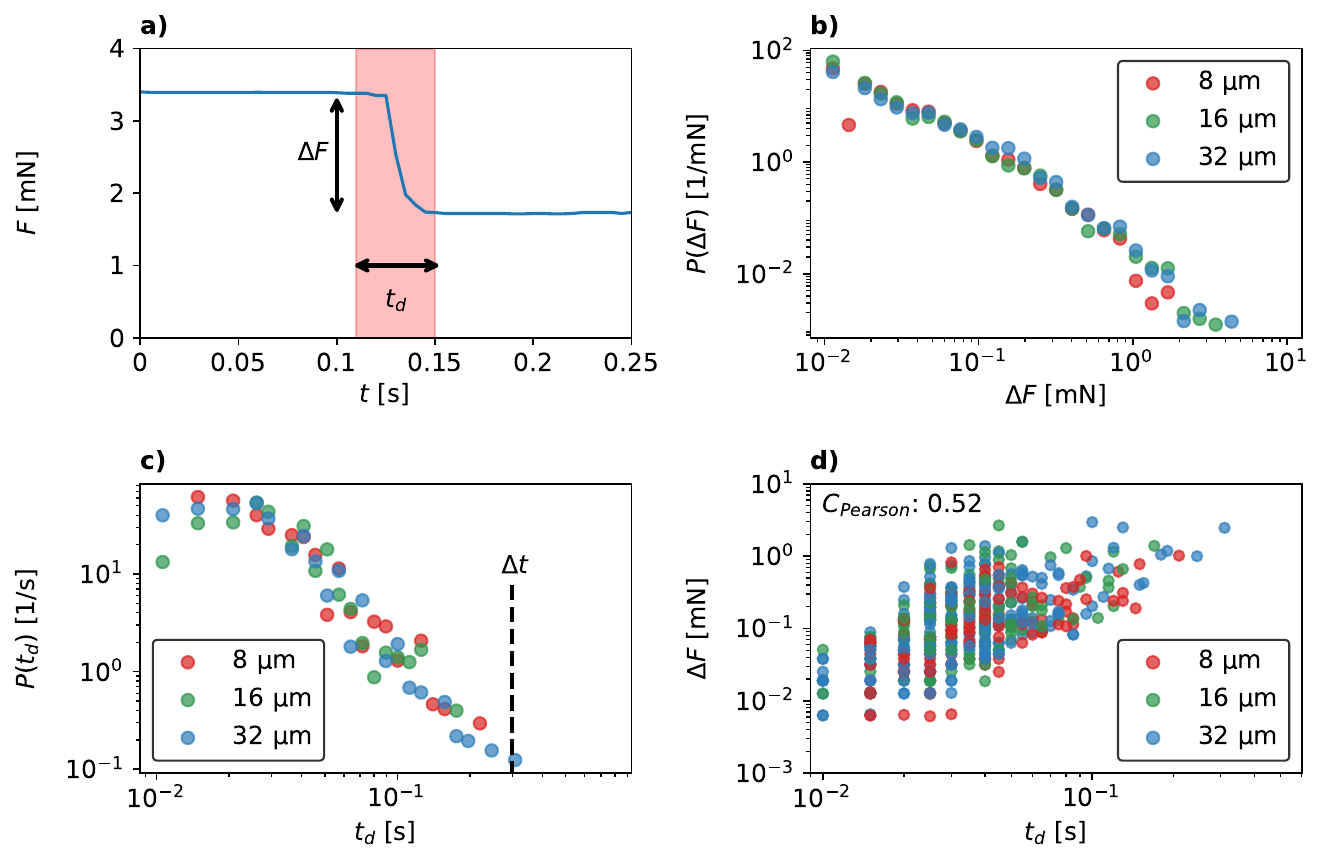}
    \caption{\textbf{Statistics of force drops and their time duration.}
    (a): Example of a force drop with drop duration $t_\mathrm{d}$.
    The probability density function of the (b) force drop and (c) drop duration for experiments of 8$\upmu\mathrm{m}$, 16$\upmu\mathrm{m}$ and 32$\upmu\mathrm{m}$ specimen sizes. 
    The chosen time window width $\Delta$t for the fine-scale prediction is indicated by the dashed line and shows that the drop durations of almost all the force drops are below the chosen time window.
    (d): Correlation between force drop and drop duration reveals an overall Pearson correlation of 0.52 (with a Pearson correlation of 0.60 for 8$\upmu\mathrm{m}$, 0.36 for 16$\upmu\mathrm{m}$ and 0.61 for 32$\upmu\mathrm{m}$).
    }   
    \label{SM:fig:duration}
\end{figure}

\begin{figure}[H]
    \centering
    \includegraphics[width=\textwidth]{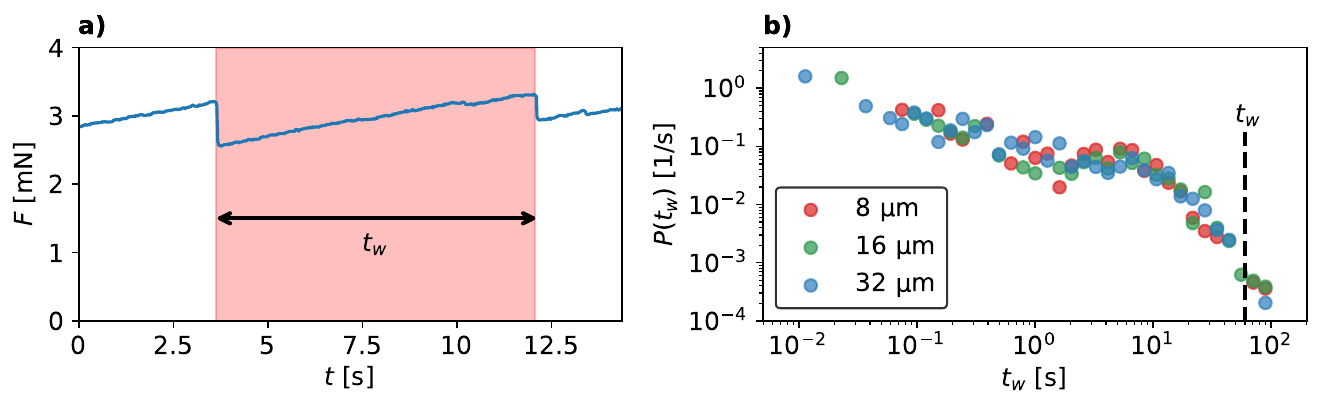}
    \caption{\textbf{Statistics of waiting times between force drops.} (a): Example of the waiting time $t_w$ between two consecutive force drops.
    (b): The probability density function of the waiting time for experiments of 8$\upmu\mathrm{m}$, 16$\upmu\mathrm{m}$ and 32$\upmu\mathrm{m}$ specimen size.
    The chosen window width $\Delta T$ for the coarse-scale prediction is indicated by the dashed line and shows that the coarse time scale is significantly larger than the typical waiting time between deformation events.
    }
    \label{SM:fig:waiting_time}
\end{figure}

\begin{figure}[H]
    \centering
    \includegraphics[width=0.5\textwidth]{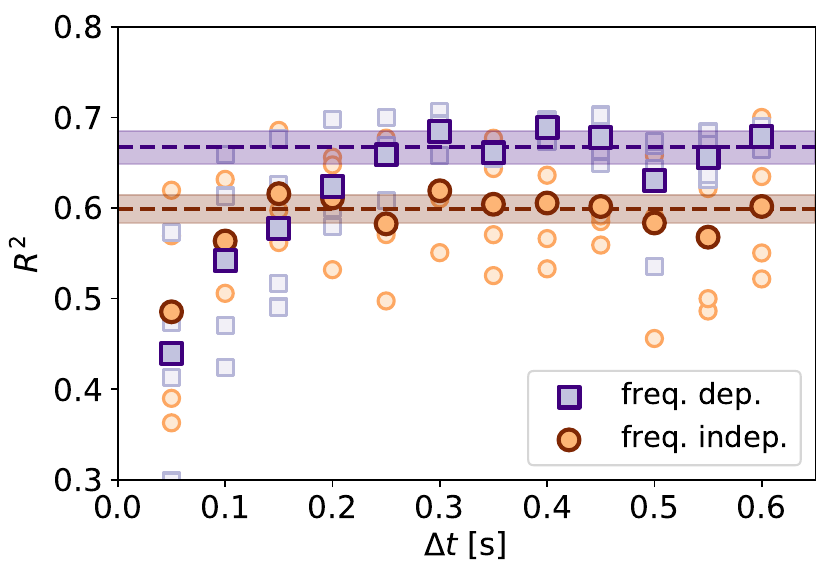}
    \caption{\textbf{The $R^2$ score for different time-window widths in case of the fine-scale
    prediction for 8-micron samples.} Our convergence study show that around 250~ms the $R^2$ saturates to $0.60\pm0.02$ and $0.65\pm0.03$ for the frequency independent and frequency dependent models, respectively. Transparent markers indicate scores corresponding to individual experiments and the larger ones indicate the mean scores. For each prediction the ML model was trained on the other three 8-micron samples.
    }
    \label{SM:fig:windows_width}
\end{figure}

\clearpage
\newpage
\section*{Supplementary note 2: The used wavelet}
\label{sec_sup:Wavelet}
\begin{figure}[H]
    \centering
    \includegraphics[width=\textwidth]{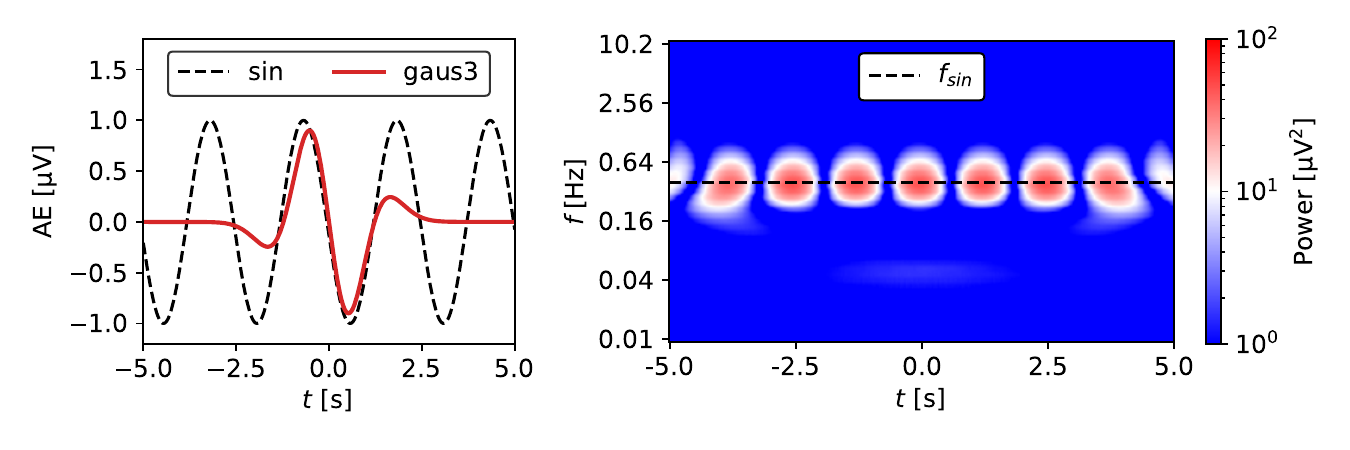}
    \caption{\textbf{Third-order Gaussian wavelet.}
    (a): The shape of the third-order Gaussian wavelet (gaus3) as well as the corresponding sinusoidal curve with an equal central frequency. (b): The wavelet transformation of the sinusoidal curve using the gaus3 wavelet.
    }
    \label{SM:fig:examplewavelet}
\end{figure}

A wavelet transformation leads to a time-frequency resolution of a time series.
A real-valued wavelet function underlies certain requirements.
For one, a wavelet function must have a zero mean~\cite{Addison2017}, and second, the wavelet energy~$E$ must be finite such that
\begin{equation}
    E = \int_{-\infty}^{\infty} |\psi(t)|^2 \, dt \, < \, \infty.
\end{equation}
A wavelet function is shown on the left side in Suppl.~Fig.~\ref{SM:fig:examplewavelet} for the chosen third-order Gaussian wavelet (gaus3).
Additionally, as a guide for the eye, the corresponding sinusoidal curve with the same  frequency as the central frequency of the wavelet function is indicated.
An overview of Gaussian wavelets in general is provided by Shao and Ma~\cite{Shao2003}.
On the right side, the corresponding time-frequency spectrogram of the sinusoidal curve based on the wavelet transformation is illustrated.
As expected, the resulting energy of the spectrogram is the largest for frequencies similar to the frequency of the sinusoidal curve.

\clearpage
\newpage
\section*{Supplementary note 3: Signal shape and power spectrum}
\label{sec_sup:Frequencies}

\begin{figure}[H]
    \centering
    \includegraphics[width=\textwidth]{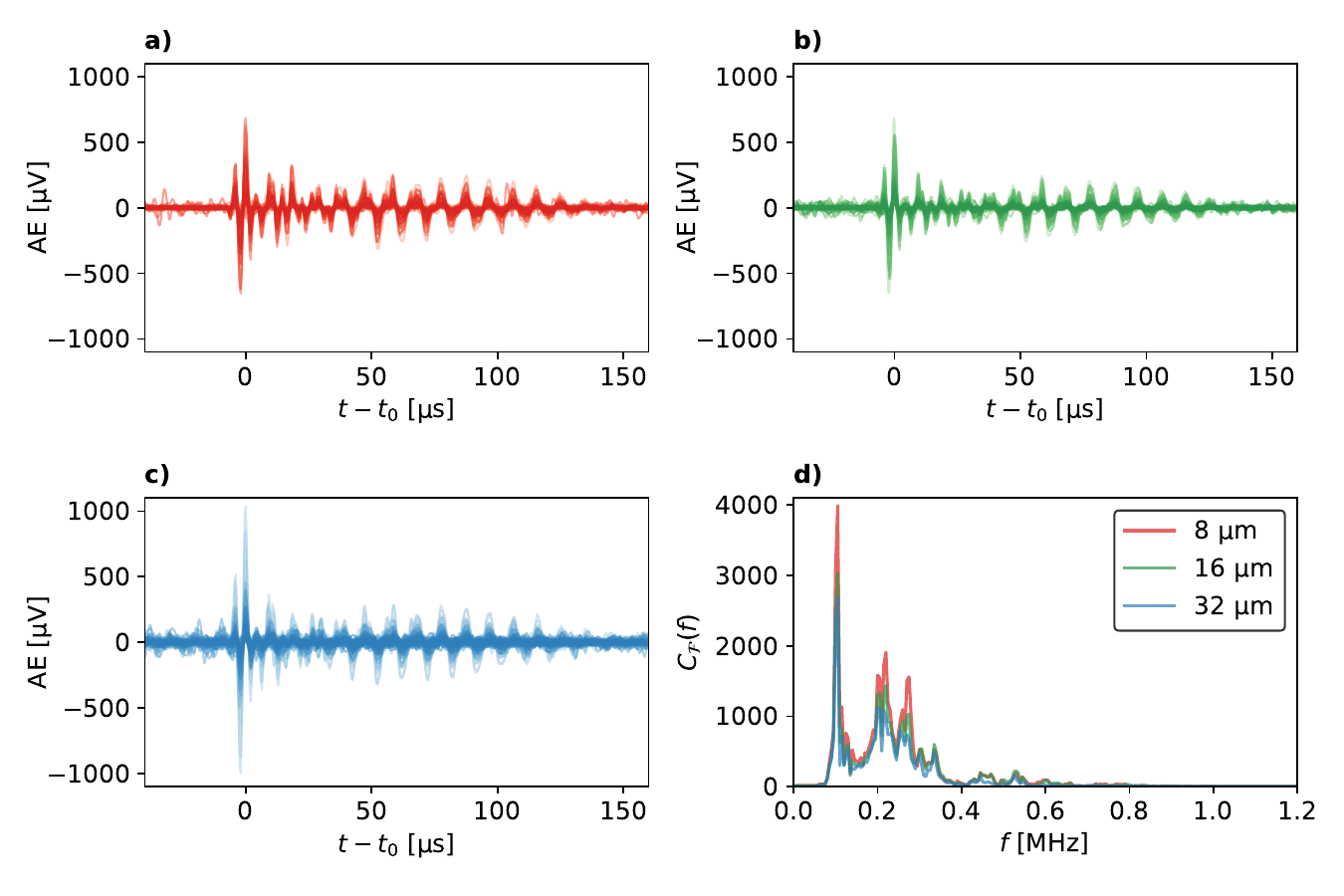}
    \caption{\textbf{Typical signal shapes.} Several typical signals
     of (a) an 8$\upmu\mathrm{m}$ (b) an 16$\upmu\mathrm{m}$ and (c) an 32$\upmu\mathrm{m}$ specimen sized experiments plotted together. $t$ is the elapsed time and $t_0$ is the onset time of the signal. Signal are translated in a manner that they start at the same transformed time. (d): The mean power spectra of the signals for the different pillar diameters. The spectrum is visibly very similar for each pillar size which allows us to apply our ML method to unseen sample sizes.
    }
    \label{SM:fig:powerspectra}
\end{figure}

\clearpage
\newpage
\section*{Supplementary note 4: Example spectrograms}
\label{sec_sup:Spectrogram}
\begin{figure}[h]
    \centering
    \includegraphics[width=\textwidth]{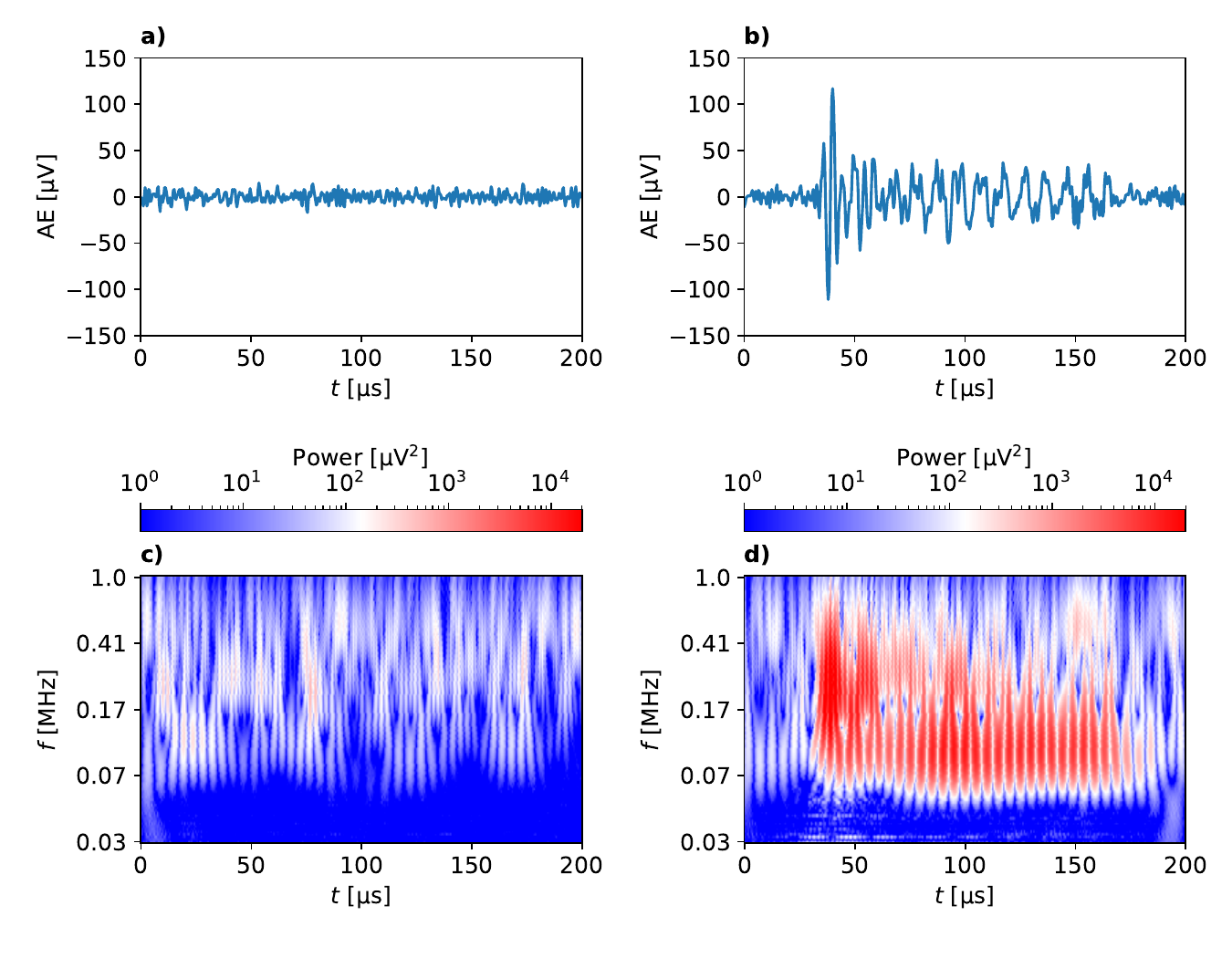}
    \caption{\textbf{Spectrograms of noise and of an acoustic signal.} AE signal (a) during a period without any detected dislocation motion and (b) during a dislocation avalanche as well as the corresponding spectrograms (c) and (d), respectively, transformed by a continuous wavelet transformation with a third-order Gaussian wavelet.}
    \label{fig_sup:Spectograms}
\end{figure}

Each spectrogram is a 2D representation of time $t$ and frequency $f$.
Supplementary~Fig.~\ref{fig_sup:Spectograms} displays an AE signal~a) during a period without any detected dislocation motion and~b) during a dislocation avalanche for a time window of $t=160 \upmu s$.
The corresponding spectrograms (with frequency range of 30~kHz to 1000~kHz) are generated by continuous wavelet transformation with a third-order Gaussian wavelet and are shown in c) and d), respectively.
No distinct pattern is visible in the spectrogram if there is no ongoing dislocation motion  c).
In comparison, the onset of a dislocation avalanche in~(d) at 40~µs exhibits an AE signal with a distinct frequency pattern at around 250~kHz.
The oscillating periodic pattern after the initial avalanche reveals a distinct frequency pattern at around 100~kHz until the oscillation ends.

\clearpage
\newpage
\section*{Supplementary note 5: Predictions for all experiments}

\begin{figure}[ht!]
    \centering
        \includegraphics[width=0.95\textwidth]{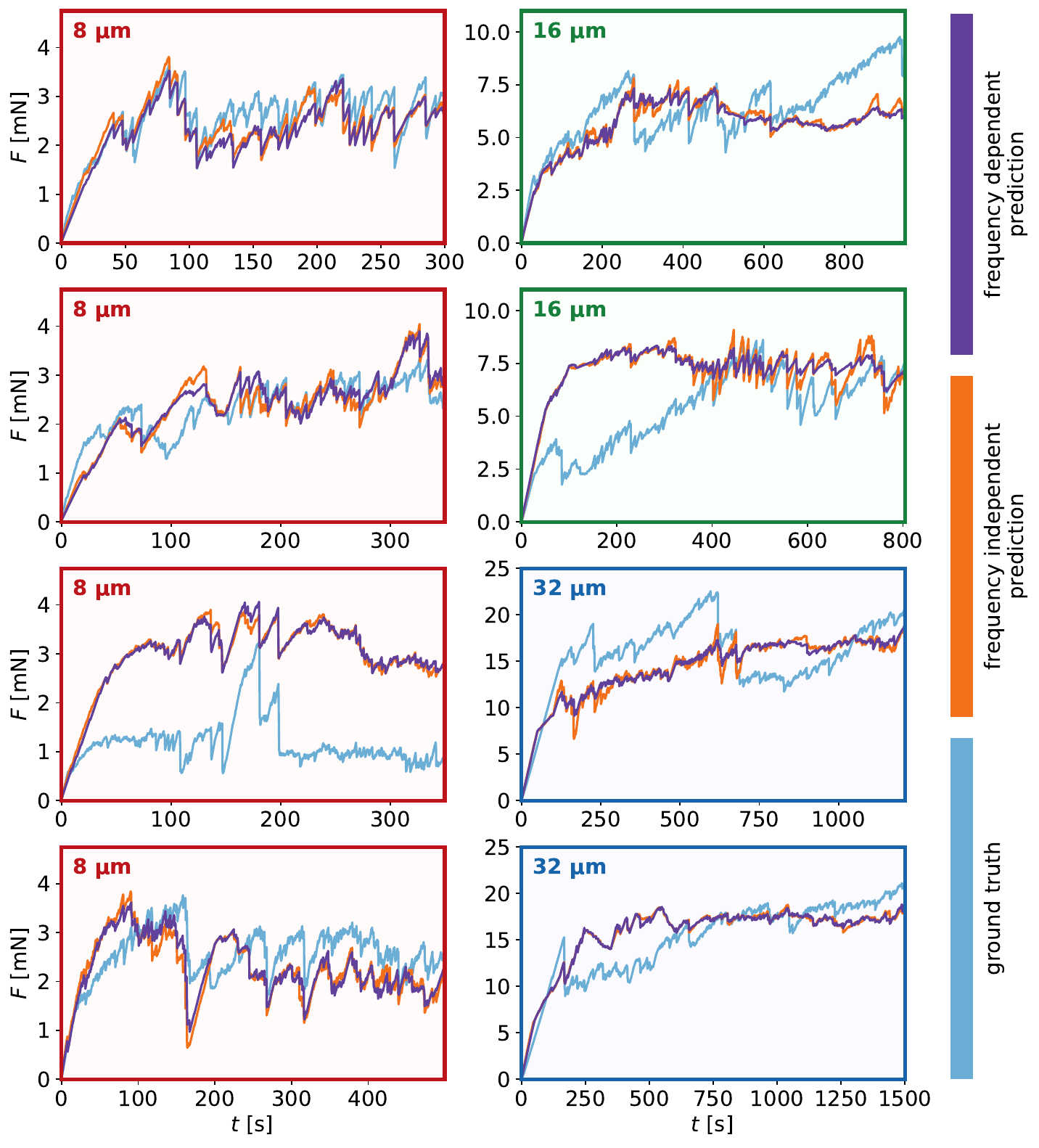}
    \caption{\textbf{All actual and predicted force-time curves.} The actual and predicted force-time ($F-t$) curves for all 8 experiments. Predictions plotted are obtained by the combination of fine-scale and coarse-scale predictions.}
    \label{fig_sup:all_pred}
\end{figure}

\clearpage
\newpage
\section*{Supplementary note 6: Size effects}

\begin{figure}[ht!]
    \centering
    \includegraphics[width=\textwidth]{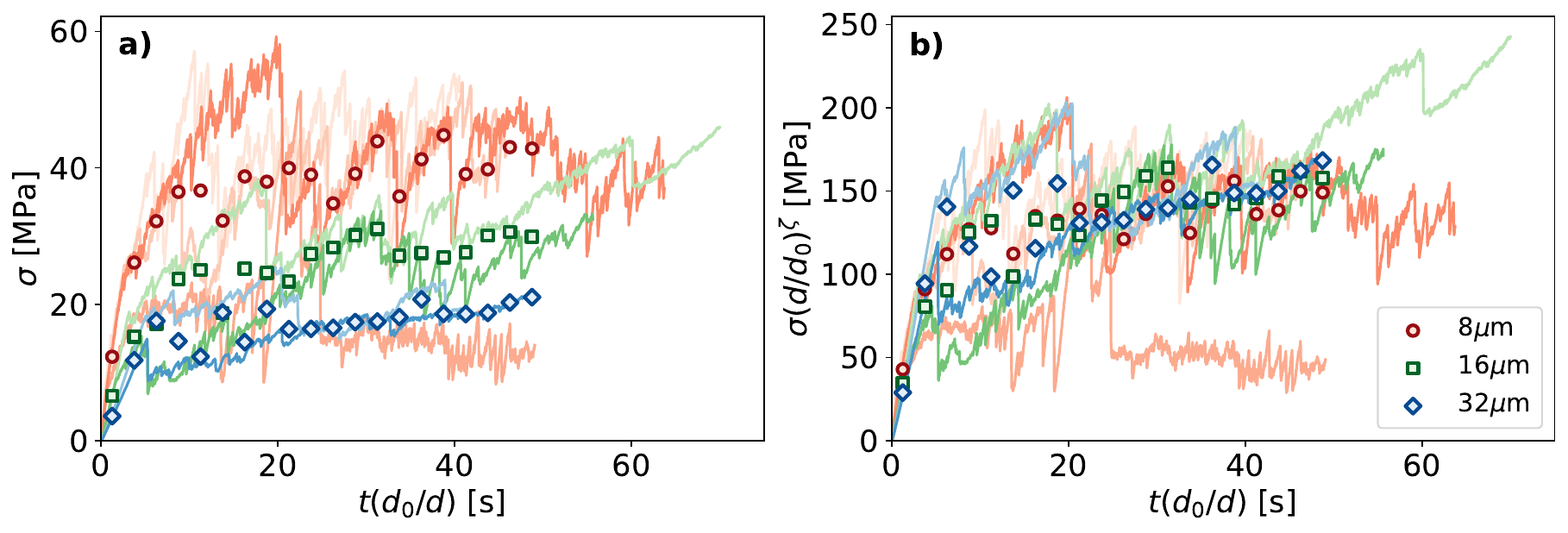}
    \caption{\textbf{Size-related hardening of micropillars.} (a):  Curves represent individual stress-time curves $\sigma$ and $t$ being stress and time, respectively. Stress is computed according to $\sigma=F/d^2$ where $F$ is the force and $d$ is the sample diameter. Rescaled time $t(d_0/d)$ is directly proportional with the strain. Here $d_0=1~\upmu\mathrm{m}$. Markers denote time-resolved median stresses. (b): Curve collapse obtained with exponent $\zeta=0.6$ showing similarity of the plastic response at different scales.}
    \label{fig_sup:size_effect}
\end{figure}

\end{document}